\documentclass{PoS}
\usepackage{url}

\newcommand{\mnras}{MNRAS~}
\newcommand{\apjl}{Astrophysical Journal Letters~}
\newcommand{\apj}{Astrophysical Journal~}
\newcommand{\aap}{Astronomy \& Astrophysics~}
\newcommand{\pasp}{PASP~}
\title{Concluding Remarks at the Multifrequency
  Behaviour of High Energy Cosmic Sources - XIII Workshop -- II}

\ShortTitle{Concluding Remarks - II}

\author{\speaker{Paolo Padovani}\\
        European Southern Observatory, Karl-Schwarzschild-Str. 
2, D-85748 Garching bei M\"unchen, Germany\\
        E-mail: \email{ppadovan@eso.org}}

\abstract{I summarise the concluding remarks I gave at the Multifrequency
  Behaviour of High Energy Cosmic Sources - XIII Workshop. That was {\it
    not} a summary talk and was meant to be provocative. I first give what
  I think the main message of the workshop was, then provide some (biased)
  highlights, touch upon the upcoming new facilities and the issues of
  ``quantity vs. quality'' and productivity in astronomy, and finally
  conclude with a look to the future. Astronomers who did not attend the
  workshop might still find the first two topics 
  appealing and the last two thought-provoking.}

\FullConference{Multifrequency Behaviour of High Energy Cosmic Sources -
  XIII - MULTIF2019\\ 3-8 June 2019\\ Palermo, Italy}

\begin{document}

\section{The end of an era}\label{sec:Intro}

If I had to summarise in one sentence this workshop, I would say this: {\it
  It marked the end of an era and the beginning of a new one.} People have
a tendency to think that their epoch is somewhat special compared to
previous ones but in our case, it really is! In the last couple of years we
have witnessed the birth of (non-stellar) multi-messenger
astrophysics\footnote{This birth happened actually in 1968 with the
  discovery of solar neutrinos and then (again) with the detection of
  neutrinos from SN 1987A. Hence the addition of ``non-stellar''.}. First,
there was the detection of the first gravitational wave (GW) event by LIGO
in 2016 \cite{Abbott_2016} [Gondek-Rosinka, Poggiani\footnote{I include the
    names of the speakers who dealt with the topic at hand during the Workshop.}], followed by
the association in 2017 between a LIGO/Virgo event (GW170817) and
electromagnetic emission from a binary neutron star merger (GRB 170817A) in
the galaxy NGC 4993 at $z=0.01$ \cite{Abbott_2017_b} [D'Avanzo]. Then there
was the first association of high-energy IceCube neutrinos with a blazar at
$z=0.3365$, TXS 0506+056, in July 2018
\cite{2018Sci...361.1378I,2018Sci...361..147I,Padovani_2018} [Righi,
  Paredes, Padovani]. And, as a bonus, this year we also had the first image of the
``shadow'' of a black hole (BH) \cite{EHT_2019} [De Laurentis]. What an
amazing time to be an astronomer!

While the astrophysical relevance of the GW170817/GRB 170817A event has been
discussed at length in the literature, I feel this is not the case for the
neutrino result. This has a number of astrophysical implications
\cite{Padovani_2018}: 1. with neutrinos we are now exploring an energy
range ($\sim$ PeV $= 10^{15}$ eV) which is, and will always be,
inaccessible with photons at this (or any!) redshift. That is because these very
high-energy photons collide with IR/mm photons (the so-called
extragalactic background light [Costamante, Paredes]) and are annihilated 
with the
resulting production of electron
-- positron pairs. Neutrinos provide us then with a new and unique window on blazar
physics; 2. the spectral energy distribution of at least one blazar has to
be modelled using protons (the so-called lepto-hadronic scenario), laying
to rest a debate (leptons or hadrons?), which has been around for decades;
3. the number of known neutrino sources has jumped by 50\% from two
(the Sun and SN 1987A) to three. And we have identified the first
non-stellar neutrino source; 4. the {\it first} cosmic ray (CR) source has
been identified. The IceCube results, in fact, imply the existence of
protons with energies $\gtrsim 1$ PeV, around the so-called "knee" of the
CR flux distribution [Caccianiga], in the blazar TXS 0506+056. And as it often
happens when new, ground-breaking observations are available, theorists have
some difficulty in explaining in a coherent way the electromagnetic and
neutrino emission in TXS 0506+056 (e.g. \cite{Keivani_2018,Winter_2019})
[Righi].

\section{Selected topics at this workshop}\label{sec:topics}

 \begin{figure}
 \center
     \includegraphics[width=1.0\textwidth]{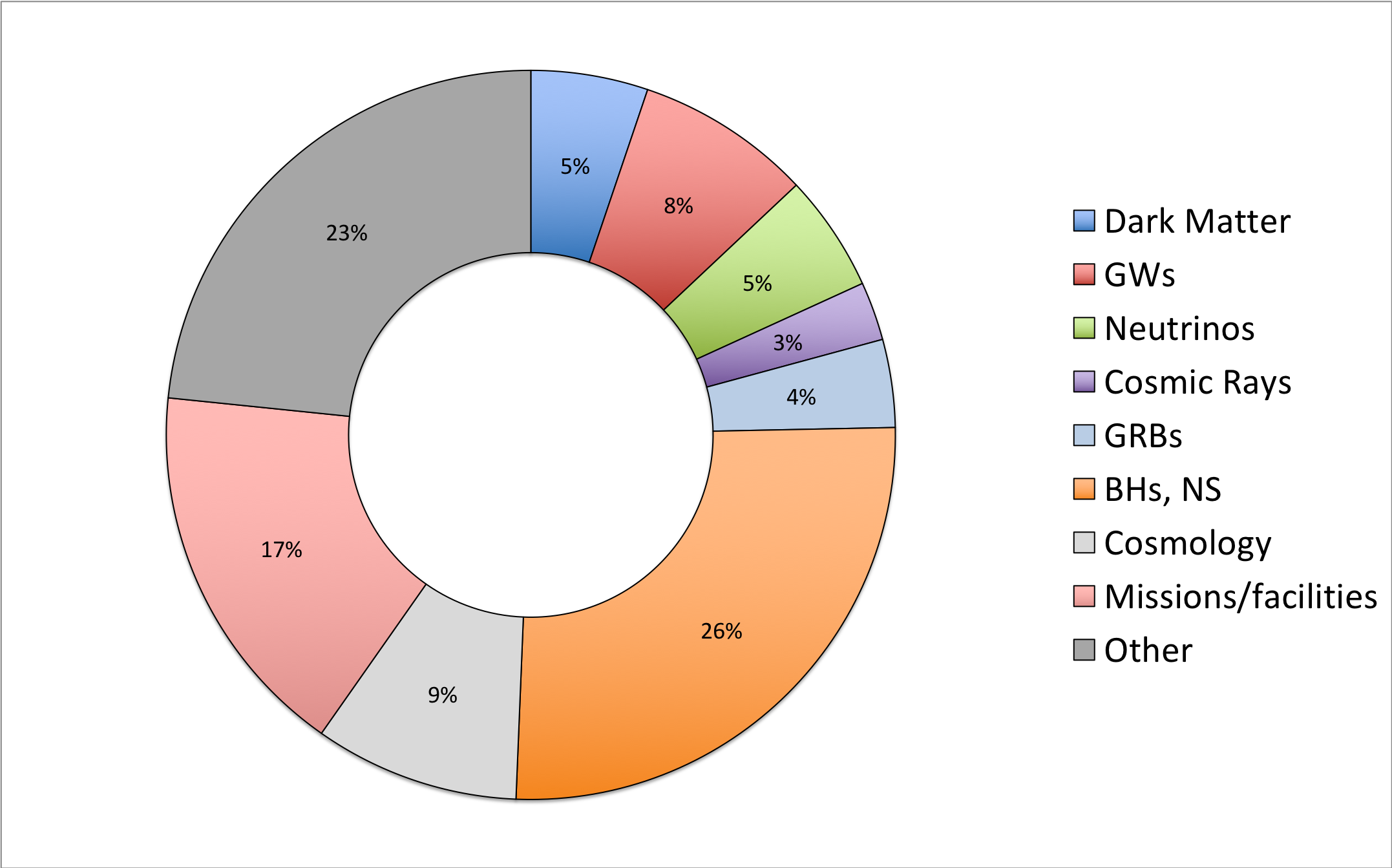}
     \caption{The distribution of talks at the workshop grouped by topics:
       21\% of the talks dealt with multi-messenger astronomy.}
     \label{fig:chart}
     \end{figure}

The advent of this ``new era'' is also apparent in the topics discussed at
the workshop, which are shown in Fig. \ref{fig:chart}: a conservative
estimate shows that 21\% of the talks dealt with multi-messenger (i.e., non
``photon-based'') astronomy. This is a considerable fraction, which reflects 
also the fact that this was the
first Frascati Workshop after the LIGO/Virgo and IceCube results. I now
touch upon some selected (and biased) topics, which were discussed at the
workshop.

\subsection{Peak luminosities of Gravitational Wave events}

The GW events reach extraordinary values of peak luminosity, so far all but
one above $10^{56}$ erg s$^{-1}$ (Tab. III of \cite{Abbott_2019}), albeit
over a very short time ($\approx$ a few ms) [Gondek-Rosinka,
  Poggiani]. These can be compared to the powers of 
  Active Galactic Nuclei (AGN), with $L \lesssim 3 \times
10^{48}$ erg s$^{-1}$ over more than tens of millions of years, and
$\gamma$-ray bursts, with $L \lesssim 5 \times 10^{54}$ erg s$^{-1}$ over a
few seconds. Amazingly enough, the GW luminosities are even larger than the
power emitted by all the stars in the Universe! An upper limit to this
value can be roughly calculated by multiplying the luminosity of the Milky
Way ($ \sim 8 \times 10^{43}$ erg s$^{-1}$) by the number of
galaxies in the Universe ($\approx 10^{12}$: \cite{Conselice_2016}), which
gives $L_{\rm all~stars} < 8 \times 10^{55}$ erg s$^{-1}$ (as most galaxies
are less luminous than the Milky Way).

These powers are also quite close to the {\it maximum} luminosity of any
physical system, which is sometimes called the Planck luminosity $L_{\rm
  Planck}$ \cite{Thorne_1983,Abbott_2017_a}.  This can be derived by
dividing the rest mass energy of a body ($M c^2$) by the crossing time of
its event horizon ($\sim [2 G M/c^2]/c$), which yields $L_{\rm Planck}/2$,
where $L_{\rm Planck} = c^5/G = 3.63 \times 10^{59}$ erg s$^{-1}$.

On a separate note, although the LIGO/Virgo data are not sensitive to the
signal of binary supermassive black hole (SMBH), GW data have already been used in this context:
\cite{Jenet_2004} have excluded the presence of a binary SMBH in 3C 66B, a
radio galaxy at $z=0.02$, using pulsar timing [Possenti]. Were this binary
BH there, in fact, as had been suggested by radio observations, it would
have been apparent in the 7 yr of timing data from the radio pulsar PSR
B1855+09. 

In general, accreting systems, ranging from white dwarf binaries to
cataclysmic variables to X-ray binaries to AGN, are potential gravitational wave
emitters at different scales [Poggiani]. The spectrum of
GWs from these systems is very broad and their detection requires ground based
interferometry, space based interferometry, and pulsar timing.

\subsection{Getting really close to black holes}\label{sec:BH_close}

The Event Horizon Telescope (EHT) has given us the first ever images of the
``shadow'' of the BH at the centre of M 87 \cite{EHT_2019} [De Laurentis],
which has allowed the EHT collaboration to determine its mass ($6.5\pm0.7
\times 10^9$ M$_{\odot}$). The diameter of the shadow is $\sim 5.5$ times
the Schwarzschild radius of the SMBH.  It is estimated
that these images have been seen by about three billion people (H. Falcke, p.c.),
which would make these the most popular images of all times. With a jet
inclination angle $\sim 17^{\circ}$ \cite{Walker_2018} M 87 is {\it almost}
a blazar of the BL Lac type \cite{Urry_1995} (see also Sect. \ref{sec:blazars}).


GRAVITY at the ESO/VLT is also getting very close to the BH at the centre
of the Milky Way by studying flares at distances $\sim 3 - 5$ times the
Schwarzschild radius, with significant and continuous positional changes of
the emission centroid corresponding to $\sim 30\%$ the speed of light
\cite{Gravity_2018} [Borkar]. The Extremely Large
Telescope\footnote{\url{http://www.eso.org/sci/facilities/eelt/}}, which in
2025 will be the largest optical-near-IR telescope in the world with a
diameter of 39m [Padovani], will be able to detect a $10^6$ M$_{\odot}$ BH up
to $\sim 30$ Mpc and a $10^9$ M$_{\odot}$ one up to $\sim 1$ Gpc. This will
provide an increase in the distances reachable with current 8-10m
telescopes of a factor $\approx 5$. The ELT, thanks to its much better
resolution, will also be able to get even closer to BHs than currently
possible.

\subsection{Black hole spins}

We have also heard about BH spins [Aschenbach, Bambi]. The mantra in
the AGN community has been for years that jetted
AGN \cite{Padovani_2017} (also called, I think misleadingly,
``radio-loud'') are powered by a rotating BH while non-jetted AGN are not
\cite{Padovani_2017_b}. X-ray reflection spectroscopy is currently the only
available method to measure the spin of SMBHs and $70\%$ of AGN have very
high $a_*$ ($> 0.9$, where $a_* = c J/G M^2$ is the dimensionless BH spin
parameter and $J$ is the angular momentum) [Bambi]. But most of these are
non-jetted AGN, which should be non-rotating. However, these spin
measurements have to be taken carefully, as they may be affected by
systematics related to the model employed to infer them \cite{Bambi_2018}.

M 87 is a classical jetted AGN. I would have then expected the EHT
collaboration (Sect. \ref{sec:BH_close}) to find a relatively high value of
$a_*$. However, the M 87 papers just say that ``compact 1.3 mm emission in M87
arises within a few $r_g$ of a Kerr BH"  \cite{EHT_2019_1}. So we know that the BH at the
centre of M 87 is rotating but more data are needed before a specific value of $a_*$ 
can be provided. This reflects the fact that the size of the
shadow of a Kerr BH depends only weakly on spin (and inclination)
\cite{Psaltis_2019}.

The LIGO/Virgo GW data can also constrain spins [Gondek-Rosinka, Poggiani],
in this case of stellar BHs.  Table III of \cite{Abbott_2019} provides
values of $\chi_{\rm eff}$, which is a mass-weighted linear combination of
the spins of the two merging BHs projected onto the Newtonian angular
momentum. Only in 2/11 cases can $\chi_{\rm eff} = 0$ be excluded at $>
90\%$ confidence, which means that the data disfavour scenarios in which
most BH merge with large spins aligned with the binary's orbital angular
momentum.

\subsection{Magnetars}

I really enjoyed learning about magnetars [Mereghetti, Nakagawa, Dainotti,
  D'Avanzo, Ferrazzoli]; see also \cite{Mereghetti_2015}. A magnetar is a
type of neutron star having a huge external magnetic field $\sim 10^{13} -
10^{15}$ G, i.e., up to $\sim 1,000$ above the average. This is the main
source of energy, instead of the rotation, accretion, nuclear reactions, or
cooling, which power the more normal neutron stars. What are now called
magnetars were initially split into two separate classes of sources, soft
$\gamma$-ray repeaters and anomalous X-ray pulsars. The main properties of
the two dozen magnetars known in the galaxy and the Magellanic Clouds are
their slow rotation periods (P $\sim 2 - 12$ s), persistent X-ray powers
($L_{\rm X} \approx 10^{34} - 10^{36}$ erg s$^{-1}$), faint optical/NIR
counterparts ($K \sim 20$), and strong variability, with powerful short
bursts in the X-rays and soft $\gamma$-rays, often reaching super-Eddington
luminosities. Three {\it giant} flares have also been observed with huge
peak powers. The most powerful one ($\approx 10^{47}$ erg s$^{-1}$) came
from SGR 1806--20 on December 27, 2004 and was mind-boggling: more than 20
satellites recorded this exceptional event, which started with a hard pulse
so intense that it saturated most detectors and significantly ionised the
Earth's upper atmosphere \cite{Mereghetti_2005}. The flare was brighter
than anything ever detected from beyond our Solar System with a fluence
$\sim 2$ erg cm$^{-2}$ ($E > 80$ keV) and lasted over a tenth of a
second. Magnetars are also possible GW sources both through their fast
rotation, which leads to deformations, and their impulsive activity.

\subsection{Blazars}\label{sec:blazars}

Blazars are AGN hosting a relativistic jet oriented at a small angle
($\lesssim 15 - 20^{\circ}$) w.r.t. the line of sight
\cite{Urry_1995,Padovani_2017}. This translates into very interesting and
somewhat extreme properties, including relativistic beaming, which makes
blazars appear orders of magnitude more powerful than they really are,
superluminal motion, and strong, non-thermal emission over the entire
electromagnetic spectrum and beyond, i.e., into neutrino territory. At the
meeting we heard the latest news about blazars [Costamante, B\"ottcher,
  Pittori]. These include the image of the ``shadow'' of the BH at the
centre of M 87, which is ``almost'' a blazar (Sect. \ref{sec:BH_close}) and
the first association of high-energy IceCube neutrinos with a blazar of the
BL Lac type at $z=0.3365$, TXS 0506+056 (Sect. \ref{sec:Intro}). Apparently
Nature loves disks and jets, as they seem to be present in a variety of
astronomical objects ranging from stars and planetary systems, X-ray
binaries, and AGN, including blazars of the ``flat spectrum radio quasar''
(FSRQ) type. In the latter, however, it looks like the power of
relativistic jets is larger than the luminosity of their accretion disks
\cite{Ghisellini_2014}. $\gamma$-ray emission in FSRQs is generally
explained as inverse Compton radiation of relativistic electrons in the jet
scattering optical-UV photons from the broad-line region (BLR), the
so-called BLR external Compton (EC) scenario. However,
\cite{Costamante_2018} have found no evidence for the expected BLR
absorption, with only 1 object out of 10 being compatible with substantial
attenuation, which essentially rules out the EC mechanism and implies that
$\gamma$-ray emission originates predominantly outside the BLR. This 
has important implications for the theoretical interpretation of the 
spectral energy distributions of blazars and it also means
that CTA should see many more FSRQs than previously thought. Finally, the
Astro-rivelatore Gamma a Immagini Leggero (AGILE) satellite, has found
three transient $\gamma$-ray sources ($E > 100$ MeV) temporally and
spatially coincident with recent high-energy neutrino IceCube events
\cite{Lucarelli_2019}. The post-trial chance probability for this to happen
is $\sim 4.7 \sigma$. One of the objects is the already known neutrino source 
TXS 0506+056 (Sect. \ref{sec:Intro}). For the other two there are no
obvious counterparts, although one of the most interesting sources is
(again) a blazar of the BL Lac type (3FGL J0627.9--1517).

\subsection{Supernovae (in the optical band)}

Last but not least, I wanted to mention two examples of synergy between
supernovae in the optical band and cosmology. We learnt that extinction in
starburst clusters is temporarily altered by type II SNe for $\sim 50 -
100$ Myr after the star formation episode, which has important implications
for extinction corrections in the early Universe [De Marchi]. And also that SNe
Ia are dimmer in ellipticals and brighter in spirals, which implies that
the supernova properties depend on their environment. Said differently, one
needs to include a host galaxy term into the Hubble diagram fit, which is
used to constrain the shape of the Universe [Pruzhinskaya].

\section{New facilities and the era of ``even bigger data''}\label{sec:facilities}

We have also heard about some of the new facilities, which will come online
in the next few years (or have been recently starting taking data) and will
be very relevant for the topics discussed at the workshop. These include,
without any claim to completeness (and in rough chronological order within a band):

\begin{itemize}

\item Radio: ASKAP, MeerKAT, e-MERLIN, APERTIF, SKA

\item IR: JWST, Tokyo Atacama Observatory, Euclid, WFIRST, SPICA

\item Optical/near-IR: Zwicky Transient Facility, LSST, ELT, GMT, TMT 

\item X-ray: eROSITA, IXPE, SVOM, eXTP, XIPE, Athena, Theseus, FORCE,
  XRISM, Colibr\`i

\item $\gamma$-ray: Large High Altitude Air Shower Observatory, CTA 

\end{itemize}

... and certainly more, including CubeSats [Bernardini, Caiazzo,
  Ferrazzoli, Hudec, Ishida, Mori, Padovani]. These are going to move us
from the ``big data'' era into the "even bigger data" era. For example,
while the volume of the Sloan Digital Sky Survey was around 40 Terabytes,
the LSST will reach 200 Petabytes, while the SKA will get into Exabyte
territory \cite{Zhang_2015}.

\subsection{Quantity vs. quality}

 \begin{figure}
    \center
     \includegraphics[width=.7\textwidth]{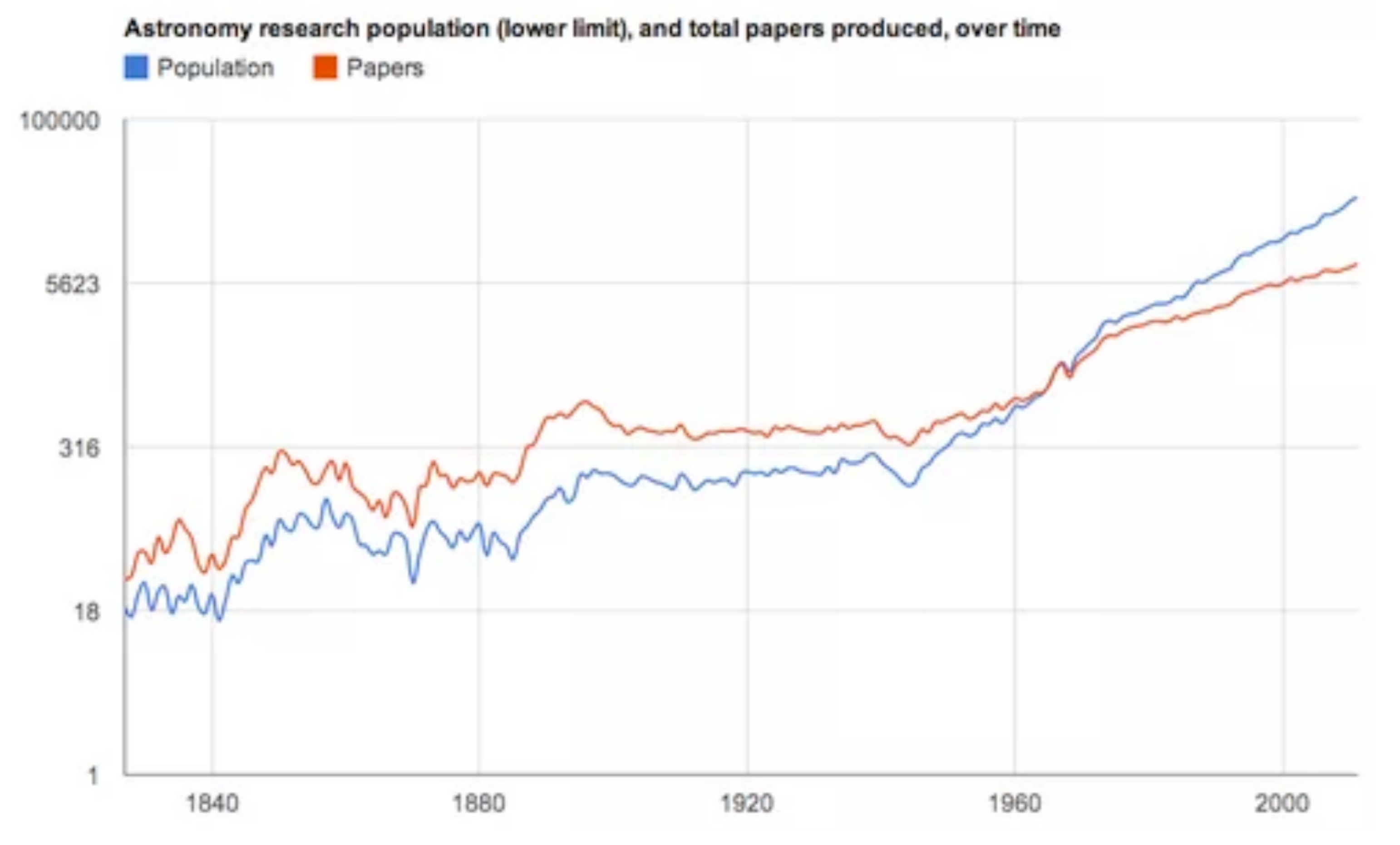}
     \caption{The number of astronomical papers (red line) and publishing
       astronomers (blue line) per year vs. year. Note the logarithmic
       scale on the y-axis. Courtesy of Robert Simpson: see
       \protect\url{https://orbitingfrog.com/2012/08/04/authorship-in-astronomy/}.}
     \label{fig:papers}
     \end{figure}

More data might mean more papers. Can we see signs of increasing 
astronomical output also in terms of published papers? Yes, as shown
in Fig. \ref{fig:papers} (red line). After a slow rise the numbers have started to
pick up after around 1960. Is this a good sign?  Well, about 50\% of all
science papers have $\leq 1$ citations, based on a study of about 58
million papers published since 1900 \cite{VanNoorden_2014}. And the
percentage of papers published in 2009 with no citations other than
self-citations after 5 years, based on 120 million academic papers, is
72.1\%, down from even higher values in previous years \cite{Fire_2019}.

The case for astronomy looks much better, although the studies I found were
old and/or limited in numbers: 3.3\% of papers (283/7724) published in 2001
and 2002 in 20 journals are never cited during the three calendar years
after the one in which they were published \cite{Trimble_2007}; and 6.1\%
of papers (20/326) published in 1961 in American journals are never cited
during the 18 years after publication \cite{Abt_1981}. To get some better
statistics I picked two random contiguous years (2000 and 2001) and used
the Astrophysics Data System. Out of the 39,972 astronomical refereed
papers published in that time period 15.8\% (6327) gathered zero citations
to date. Most of these are in (many) minor journals. If I consider only AAS
journals, A\&A and A\&AS, MNRAS, PASP, and PASJ the fraction of never cited
papers gets much
smaller, i.e., 1.3\% (159/12568).

But is quantity correlated with quality? Or is there actually an inverse
correlation? A ``destructive feedback between the production of
poor-quality science, the responsibility to cite previous work and the
compulsion to publish'' has been pointed out \cite{Sarewitz_2016} (see also
Franco Giovannelli's introductory remarks at this workshop). The same paper
talks also about the fact that ``Current trajectories threaten science with
drowning in the noise of its own rising productivity ... Avoiding this
destiny will, in part, require much more selective publication. Rising
quality can thus emerge from declining scientific efficiency and
productivity. We can start by publishing less, and less often''. Granted,
this paper deals with biology but some of these points apply to astronomy
as well. Whose fault is it? At least partly ours. We as referees, together with the 
journal editors, allow way too many papers
to get published\footnote{I am doing my best to go against the tide: in the
  past four years I have rejected 73\% of the papers I have refereed. I simply
  require that papers add something {\it substantially new} to our understanding of the
  topic at hand.}. Moreover, I have the feeling that the system, to which we all belong, tends to give way too 
  much importance to quantity, which is easier to evaluate, and less to quality.

 \begin{figure}
     \center
     \includegraphics[width=.7\textwidth]{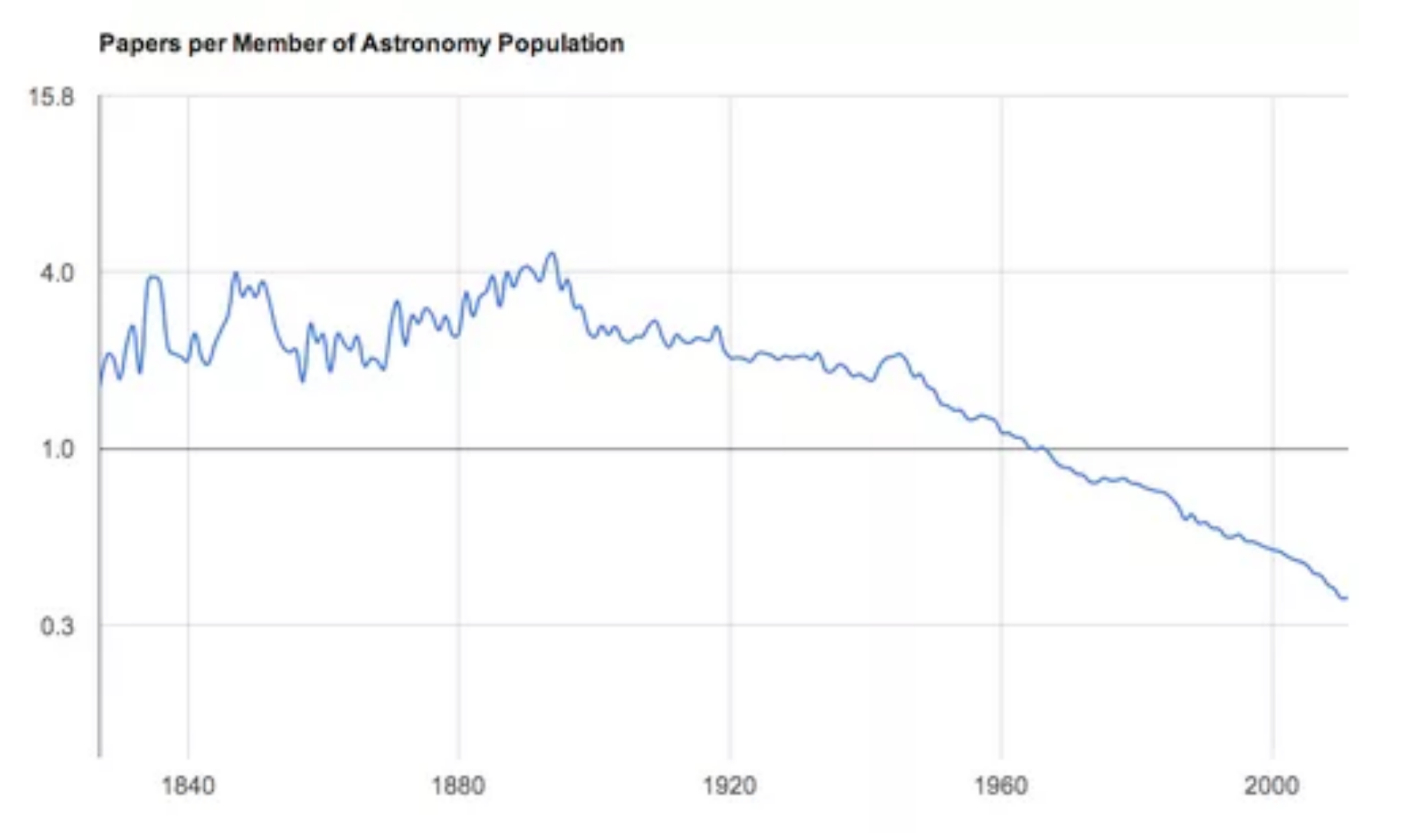}
     \caption{The number of astronomical papers per capita per year
       vs. year. This is the ratio of the red and blue lines in Fig. \ref{fig:papers}. 
       Note the logarithmic scale on the y-axis. Courtesy of
       Robert Simpson: see
       \protect\url{https://orbitingfrog.com/2012/08/04/authorship-in-astronomy/}.}
     \label{fig:papers_per_author}
     \end{figure}

But are we really publishing more? No: per capita we are publishing less!
Fig. \ref{fig:papers_per_author} shows that while in 1960 astronomers were
publishing about 1 paper each per year, we are now getting close to 1 paper
every three years. Every colleague I show this figure to is shocked, as we
are all under the impression that we are publishing a lot. But the number
of papers has not caught up with the increasing number of astronomers,
which means we are getting less efficient. My interpretation of
  this result has to do with the increasing size of astronomical
  collaborations: while a group of $2-4$ astronomers might easily publish
  $2-4$ papers per year, a large collaboration of, say, 100 people, is {\it
    not} going to publish 100 papers per year. One might however argue that
  papers in 1960 were shorter \cite{Abt_2000} and therefore easier to
  write, although I am not sure this effect by itself can explain the trend shown in Fig. 
  \ref{fig:papers_per_author}.

\section{Getting ready}\label{sec:ready}

Let us look at the bright side: we will soon be (even more) flooded with
data relevant to the topics discussed at this workshop. And we need to be
ready for that. What follows is some advice (mostly for the younger members
of the astronomical community):

\begin{itemize}

\item Think out of the box: even now there are lots of data but very few
  new ideas.  Spend less time running around writing papers and more
  time thinking about the important open issues.

\item Ask the right questions: doing that is the toughest part of solving
  problems.

\item Change topic every once in a while. I love this quote from Pablo
  Picasso: ``Success is dangerous. One begins to copy oneself, and to copy
  oneself is more dangerous than to copy others. It leads to sterility.''
  Astronomy is great also because one can change band and topic relatively
  easily. It takes some courage and humility, as you need to start from
  scratch in a new field, but it is very rewarding.

\item Learn the right tools. Be they ``data mining'', ``virtual
  observatory'', ``neural networks'', ``artificial intelligence'',
  whatever. It is clear that we cannot handle the huge amount of data we
  will soon get without changing the way we deal with them. As a first
  step, have a look at the presentations given at the Workshop on ``Artificial
  Intelligence in Astronomy'' held at ESO in June
  2019 ({\url{https://www.eso.org/sci/meetings/2019/AIA2019.html}}).
  
\item Have fun! This is the most important advice. If you do not enjoy what
  you are doing you will be much less productive and also less happy.

\end{itemize}

I want to conclude with a plea to Franco Giovannelli to change the workshop
name to {\it Multimessenger Behaviour of High Energy Cosmic Sources}!

\acknowledgments

I thank the organisers of the workshop for their kind invitation, Evanthia
Hatziminaoglou, Elisa Resconi, and Eva Villaver for their careful reading
of the paper, and Mariafelicia De Laurentis, Sandro Mereghetti, and Rosa
Poggiani for useful discussions.

\end{document}